\documentclass[prd,aps,showpacs,tightenlines,superscriptaddress,amsmath,amssymb,amsfonts,twocolumn]{revtex4}
\usepackage{amssymb,amsmath,amsthm,amsbsy,epsfig,color,graphicx,times,bbm}
\usepackage[ansinew]{inputenc}
\usepackage[english]{babel}
\usepackage{accents}
\usepackage{color}
\usepackage{braket}
\usepackage{booktabs}
\usepackage{tabularx}
\usepackage{array}
\usepackage{bbold}

\begin{document}

\title{Gravity, entanglement and CPT-symmetry violation in particle mixing}

\author{K. Simonov}
\email{kyrylo.simonov@univie.ac.at}
\affiliation{Fakult\"{a}t f\"{u}r Mathematik, Universit\"{a}t Wien, Oskar-Morgenstern-Platz 1, 1090 Vienna, Austria}
\author{A. Capolupo}
\email{capolupo@sa.infn.it}
\affiliation{Dipartimento di Fisica ``E.R. Caianiello'' Universit\`{a} di Salerno, and INFN --- Gruppo Collegato di Salerno, Via Giovanni Paolo II, 132, 84084 Fisciano (SA), Italy}
\author{S. M. Giampaolo}
\email{sgiampa@irb.hr}
\affiliation{Institut Ru\dj er Bo\v{s}kovi\'c, Bijeni\v{c}ka cesta 54, 10000 Zagreb, Croatia}

\begin{abstract}
We study the probability oscillations of mixed particles in the presence of self-gravitational interaction. We show a breaking of the CPT-symmetry due to the contemporary violation of the T-symmetry and the CP-symmetry preservation. This violation is directly associated to the rising of the entanglement among the elements of the system that can be seen as a pure many-body effect scaling with the number of the elements in the system. This effect could have played a relevant role in the first stages of the Universe or in core of very dense systems. Experiments based on Rydberg atoms confined in microtraps can simulate the mixing and the mutual interaction and could allow to test the mechanism here presented.
\end{abstract}

\maketitle

\section{introduction}

Particle mixing and oscillations have provided some of the most direct and robust indications of physics beyond the standard model~\cite{Wu1957,Giunti2007,Griffiths2008}. We have several examples of such phenomenon both in the bosonic and fermionic sectors. In the first sector we have mixing among axion--photon~\cite{Raffelt1996,Sikivie2011,Capolupo2019_1}, $\eta$--$\eta'$~\cite{Pham2015}, neutral kaons~\cite{Christenson1964} and $B$ meson~\cite{Abashian2001}. In the second one we can find the neutrino flavor oscillations~\cite{Bilenky1978,Gonzalez2008}, the neutron--antineutron oscillations that could be observed in the next generation of experiments using slow neutrons with kinetic energies of a few $meV$~\cite{Phillips2016}, and the quark mixing~\cite{Nakamura2010}. Apart from the last one which involves particles confined inside hadrons, all the other mixing phenomena concern only neutral particles. All of them are characterized by the fact that the physical fields, called flavor fields, are superpositions of free fields with definite different masses. 

Since the difference between the masses is very small, also weak perturbations can produce measurable deviations from vacuum oscillation frequencies. The extreme sensibility to a wide set of perturbations is at the basis of different experiment proposals. For instance, many studies in recent years have been devoted to the possibility to test the quantum nature of gravity using concepts of quantum information theory~\cite{Bose2017,Marletto2017}. These proposals are based on the idea of using a system in which the intensity of the gravitational interaction depends on some internal degrees of freedom~\cite{Giampaolo2018}. The sensibility of the oscillation frequency of neutral particles such as neutrino provides a natural system where to analyze these effects~\cite{Marletto2018}. Indeed, it is well-known that neutrinos interact exclusively via gravity and weak interaction. This last interaction is stronger than gravity but has an extremely short range~\cite{Greiner2009} (about $d = 10^{-16}/10^{-18}m$) and, hence, it can be neglected for distances bigger than $d$ for which gravity survives.

The effects of gravity on the oscillation of neutral particles are not limited to a change in the frequency of flavor oscillations. Gravity is also considered as one of the possible sources of decoherence in flavor oscillation~\cite{Ellis1984,Amelino-Camelia2000,Rovelli2004} that leads to many interesting effects like the $CPT$-symmetry violation in particle mixing~\cite{Gago2001,Guzzo2016,Lisi2000,Benatti2000,Benatti2001,Capolupo2018,Capolupo2019}. In all these papers the non--unitary evolution is introduced by considering a dissipator that generates a completely positive dynamics~\cite{Benatti1997,Benatti1998}. This dissipator summarizes the effects of all possible sources of decoherence and does not allow to analyze the origin and the relative weight of the different sources of decoherence.

In the present paper we adopt a different approach to analyze the role of gravity in the particle mixing phenomena. Instead to consider a single particle as an open system affected by several uncontrolled phenomena, we consider an ensemble of self-interacting particles as a closed system where all internal physical quantities are under control. Therefore, we consider a system of $N$ mixed neutral particles evolving under the self-gravity and neglect all other possible interactions acting between the system and the environment. We prove that, because of the difference in mass of the free fields, the self-gravity induces a violation of the $T$-symmetry whereas the $CP$-symmetry is preserved. This implies a violation of the $CPT$-symmetry. Such a violation represents an emerging many-body effect associated to the rising of a non-vanishing entanglement among the different fields.

Both the entanglement and the $CPT$ violation are extremely small and difficult to be observed in laboratories. However, being many-body effects, they are related to the number $N$ of elements of the system and to the density $n$ of the particles. Therefore, the $CPT$ violation induced by the gravity could play an important role in very dense astrophysical objects and it could have affected the early stages of the Universe~\cite{Hamada2016}.

The paper is organized as follows. In Sec.~II, for the reader's convenience, we resume the main results of the quantum mechanical approach to the particle mixing. In Sec.~III we consider the effects of gravitational interaction in a simple model made of two interacting fields, and show the $CPT$-symmetry breaking and the role played by the entanglement. In Sec.~IV we generalize the formalism to the case of many interacting particles. In Sec.~V we discuss how to generalize our results to the case in which the distance among the particles changes in time and in Sec.~VI we draw our conclusions.
 
\section{Neutral particle oscillation in vacuum}

For sake of completeness, in this section we review some aspects of neutral particle oscillations. We consider the very general case represented by the mixing of two flavor fields named $n_A$ and $n_B$. To fix the ideas, in case of neutrino oscillations, the two flavor fields coincide with the leptonic flavors as $n_A = \nu_e$ and $n_B = \nu_\mu$ while, in the case of neutron--antineutron oscillations we have $n_A = n$ and $n_B = \bar{n}$. Since the flavor fields do not coincide with those of definite masses, the mixing relations can be written as
\begin{eqnarray}
\label{leptonic_state}
\ket{n_A} & = & \cos(\theta) \ket{m_1}+ e^{\imath \phi } \sin(\theta) \ket{m_2} \ \nonumber \,;\\
\ket{n_B} & = & - e^{-\imath \phi } \sin(\theta) \ket{m_1}+ \cos(\theta) \ket{m_2} \, ,
\end{eqnarray}
where $\theta$ is the mixing angle and $\phi$ is the Majorana phase which is zero in case of Dirac fermions~\cite{Giunti2007} and $\ket{m_i}$ are the states with definite masses $m_i$.

Neglecting any interaction with the rest of the Universe, the particle is a closed system which travels through space with its energy $E$. Assuming that the masses $m_i$ are much smaller than $E$, we can write the Hamiltonian of mixed fields as
\begin{eqnarray}
\label{Hamiltoniana_singolo_1}
H^{(1)}=E+\frac{c^2}{2E}(m_1^2\ket{m_1}\!\!\bra{m_1}+m_2^2\ket{m_2}\!\!\bra{m_2})\, .
\end{eqnarray}
Introducing the Pauli operator $\sigma^z$ which discriminates between the mass eigenstates, $\sigma^z\!=\!\ket{m_1}\!\!\bra{m_1}\!-\!\ket{m_2}\!\!\bra{m_2}$, and neglecting state-independent terms proportional to the identity operator in the Hamiltonian, eq.~(\ref{Hamiltoniana_singolo_1}) becomes
\begin{eqnarray}
\label{Hamiltoniana_singolo_2}
H^{(1)}=\omega_0 \, \sigma^z \; ; \;\;\;\;\;\;\;\;\; \omega_0=\frac{c^2}{4 E}(m_1^2-m_2^2) \, .
\end{eqnarray}

The single-particle state, that at $t=0$ is in one of the two flavor states of eq.~(\ref{leptonic_state}), for $t>0$ evolves under the effect of $H^{(1)}$. Due to this evolution we have, for $t>0$, a non-vanishing probability to observe a change in the flavor state of the particle. The time-dependent expression of the flavor oscillation formula is than given by $P_{ n_A \rightarrow n_B } =|\bra{n_B}\exp(-\imath H t)\ket{n_A}|^2 $ and is invariant under the exchange of flavors states, i.e. $P_{ n_A \rightarrow n_B } = P_{ n_B \rightarrow n_A }$. Explicitly, we have
\begin{equation}
\label{probability_non_interacting}
P_{ n_A \rightarrow n_B } = P_{ n_B \rightarrow n_A }=\sin^2(2 \theta) \sin^2(\omega_0 t) \,.
\end{equation}
This is the well-known Pontecorvo formula~\cite{Bilenky1978} that describes the oscillation of a neutral particle in the vacuum.

\section{Oscillation of two interacting neutral particles}

Let us now consider the case in which the system that we analyze is composed not of a single particle but of two mixing particles interacting gravitationally.

Before starting our analysis, let us discuss the basic hypothesis we will use along the section. At first, we assume the validity of the equivalence principle between inertial and gravitational mass. Moreover, we represent the gravitational interaction with the standard Newtonian potential. Replacing it with ghost-free theories of gravity as the one showed in Ref.~\cite{Biswas2012} would induce quantitative but not qualitative changes in the physical behavior. Furthermore, for the sake of simplicity, we assume that the particles travel in space with the same energy along the same direction, hence, keeping their relative distance, that we denote with $d$, a time-independent parameter. It is worth to note that this last assumption is made only to simplify the explanation of our results. Indeed, as we will show in Sec.~V, it is possible to extend our results to the case of time-dependent $d$.

Within the above assumptions the Hamiltonian of the system made of two mixed particles that interact gravitationally becomes 
\begin{eqnarray}
\label{Hamiltoniana_coppia_1}
H^{(2)}\!&\!=\!&\!H_i^{(1)}+H_j^{(1)}-\frac{G m_1^2}{d}\ket{m_1,m_1}\!\!\bra{m_1,m_1} \nonumber \\
\!&\!\! &\! -\frac{G m_1 m_2}{d}(\ket{m_1,m_2}\!\!\bra{m_1,m_2}+\ket{m_2,m_1}\!\!\bra{m_2,m_1})\nonumber \\
\!&\!\! &\! -\frac{G m_2^2}{d} \ket{m_2,m_2}\!\!\bra{m_2,m_2} \,,
\end{eqnarray}
where $G$ is the gravitational constant, $d$ the distance between the two particles and the indices in the two particles states refer, respectively, to the $i$-th particle (the first) and to the $j$-th particle (the second).

It is useful to rewrite the Hamiltonian in eq.~(\ref{Hamiltoniana_coppia_1}) in a more compact form. By recalling the definition of $\sigma^z_i$ and neglecting all terms proportional to the identity, the Hamiltonian of the system can be written as
\begin{eqnarray}
\label{Hamiltoniana_coppia}
H^{(2)}=\omega \, ( \sigma^z_i + \sigma^z_j)+ \Omega \, \sigma^z_i\cdot\sigma^z_j \, ,
\end{eqnarray}
where $\omega=\omega_0+g(m_1^2-m_2^2)$, $\Omega=g(m_1-m_2)^2$, and \mbox{$g=-\frac{G}{4d}$}. Comparing the above Hamiltonian with the one in eq.~(\ref{Hamiltoniana_singolo_2}) we can see that the presence of the gravitational interaction has two different effects. The first is that the interaction changes the value of $\omega$ from $\omega_0$ to $\omega_0+g(m_1^2-m_2^2)$ while the second one is the appearance of a new term, with amplitude equal to \mbox{$\Omega=g(m_1-m_2)^2$}, involving operators defined on both the fields. It is worth to note that both of them will disappear in the case in which $m_1$ and $m_2$ coincide. 

Before going further, let us recall two basic results of quantum information that can be found in all quantum information books as, for example, that in Ref.~\cite{Nielsen2011}. The first result is that, given a bipartite system, the evolution induced by Hamiltonian terms acting only on a single part of the system can {\it never} affect the value of the entanglement between the two parts. On the contrary, terms that act simultaneously on both the parts usually modify the entanglement. The second result is that when we project a pure state defined on a bipartite system into one of its parts, the projection obtained in such a way is still pure {\it if and only if} the state was not entangled. 

In our case, the natural bipartition of the system under analysis is obtained by considering each part coinciding with one of the two particles. With respect to this partition the terms proportional to $\omega$ are local, since each one of them acts on one single particle. Therefore, they cannot create, or destroy, entanglement inside the system. On the contrary, the term proportional to $\Omega$, induced by the presence of the state-dependent gravitational interaction, is non-local respect to the natural bipartition. Hence, it can increase, or decrease, the value of the entanglement~\cite{Wang2002}. 

At $t\!=\!0$, i.e. when the two particles were created, we can assume that there is no entanglement between them. Thus, the initial state of the whole system is a two--body fully separable (i.e. without any entanglement between the two particles) pure flavor state of the form \mbox{$\ket{\psi (0)}=\ket{n_\eta}_1\ket{n_\chi}_2$}, where $\eta$ and $\chi$ could assume all possible combinations of $A$ and $B$.

Once fixed $\ket{\psi (0)}$, the state at $t>0$ can be obtained as $\ket{\psi(t)}=U(t)\ket{\psi(0)}$, where the time evolution operator is $U(t)=\exp(-\imath t H^{(2)})$. The operator $U(t)$ is unitary because we assume that the system under analysis is closed, i.e. does not interact with the surrounding world. Therefore, for any time $t\ge 0$, the state $\ket{\psi(t)}$ is still a pure state exactly as at $t=0$. But, if $\Omega\neq0$ the state $\ket{\psi(t)}$ holds, in general, a non-vanishing entanglement between the two particles. This implies that the projection of $\ket{\psi(t)}$ on any of the two particles would be, in general, a mixed state.

It is possible to quantify how much a projection is pure using a quantity called purity defined as \mbox{$\mathcal{P}(\rho_i(t))=\mathrm{Tr}(\rho^2_i(t))$}, where \mbox{$\rho_i(t)=\mathrm{Tr}_j(\ket{\psi(t)}\bra{\psi(t)})$} is the projection over the $i$-th particle of the state $\ket{\psi(t)}$~\cite{Nielsen2011}. The purity holds a relevant role in the theory of entanglement for pure states defined in bipartite systems. Indeed, it is also associated to the $2$--Renyi entropy, defined as $S_2=-\ln(\mathcal{P}(\rho_i(t)))$, that represents a proper measure of the entanglement between a particle and the rest of the system~\cite{Horodecki2002,Plenio2007,Giampaolo2013}. Projection of fully separable states holds a purity equal to 1 and a vanishing Renyi entropy while entangled states are characterized by $\mathcal{P}(\rho_i(t))<1$ and a non-vanishing entropy. 

In our case it is easy to verify that the time-dependent expression of purity is, independently on the initial state, equal to
\begin{equation}
\label{purity}
\mathcal{P}(\rho_i(t))=1-\frac{1}{2}\sin^4(2 \theta) \sin^2(2 t \Omega) \, .
\end{equation}
In the presence of flavor mixing, i.e. for $\theta\neq n \frac{\pi}{2}$, and for any time $t \neq n \frac{\pi}{2 \Omega}$ (with $n$ integer), we have $\mathcal{P}(\rho_i(t))<1$ and, thus, the single particle state is not a pure state, hence, implying that $\ket{\psi(t)}$ is entangled. The expression of purity is also a proof of the fact that the entanglement between two flavor fields is a direct consequence of the presence of the gravitational interaction and vanishes if $\Omega$ is neglected, i.e. if $m_1=m_2$. This fact can be also considered a proof of the quantum nature of gravity since quantum correlations and entanglement, can be created only through a quantum channel~\cite{Bennet1999}.

Let us now show that, in the system we are analyzing, the presence of the entanglement induces also a violation of $T$- and $CPT$-symmetry. In order to provide this proof, we consider a simple conceptual experiment. We take into account two copies of the system already described. We assume that the two copies are identical except for the fact that, in the initial state of the first, both particles are in the state $\ket{n_A}$, i.e. $\ket{\psi_0}=\ket{n_A}\ket{n_A}$ while in the second one are both in $\ket{n_B}$ and, hence, \mbox{$\ket{\psi_0}=\ket{n_B}\ket{n_B}$}. At the same time $t>0$, we observe, in both copies, one of the two particles and analyze the oscillation probabilities.

Let us name $\rho_A(t)$ the projection obtained in the first copy of our system and $\rho_B(t)$ the one obtained in the second copy. In these two cases, the probability of flavor transitions are given, respectively, by $P_{n_A \rightarrow n_B}=\mathrm{Tr}(\rho_A(t) \ket{n_B}\!\!\bra{n_B})$ and $P_{n_B \rightarrow n_A}=\mathrm{Tr}(\rho_B(t) \ket{n_A}\!\!\bra{n_A})$. Explicitly, we obtain
\begin{eqnarray}
\label{probability_interacting}
P_{n_A \rightarrow n_B}& = & \frac{1}{2} \sin^2(2 \theta)[1-\cos(2\omega t)\cos(2 \Omega t) \nonumber \\
& & \;\;\;\;\;\;\;\;\;\;\;\;\;\;\;\;\;\;\;\; +\cos(2\theta)\sin(2 \omega t) \sin(2\Omega t) ] \; ;\nonumber \\
P_{n_B \rightarrow n_A}& = & \frac{1}{2} \sin^2(2 \theta)[1-\cos(2\omega t)\cos(2 \Omega t) \\
& & \;\;\;\;\;\;\;\;\;\;\;\;\;\;\;\;\;\;\;\; - \cos(2\theta)\sin(2 \omega t) \sin(2\Omega t) ] \; .\nonumber
\end{eqnarray}
It is easy to note that the probabilities in eqs.~(\ref{probability_interacting}) are independent on the Majorana phase $\phi$, and, hence, the $CP$-symmetry is preserved, $\Delta_{CP} = 0$. On the contrary, since the probability is not invariant under the exchange of the two flavors, we have a violation of the time-reversal symmetry
\begin{eqnarray}
\label{probability_interacting_2}\nonumber
\Delta_{T} & = & P_{n_A \rightarrow n_B} - P_{n_B \rightarrow n_A}
\\ & = &
\sin^2(2 \theta) \cos(2\theta)\sin(2 \omega t) \sin(2\Omega t) \, .
\end{eqnarray}
For $m_1\neq m_2$ we have $\Delta_{T} = 0$ only if $t = \frac{k\pi}{2\Omega}$, or $\theta=\frac{k\pi}{4},\, k\in \mathbb{Z}$. Since $\Delta _{T} \neq \Delta_{CP}$ we also have the violation of the $CPT$-symmetry. Therefore, the entanglement between the two particles induces a $CPT$-symmetry breaking. 

It is worth to note that, even if in the system that we have analyzed, the presence of the entanglement induces a violation of the $T$--symmetry, this is not a general result. Indeed, it is possible to find several Hamiltonians that can induce entanglement without breaking the $T$-symmetry, or breaking both the $T$- and the $CP$-symmetry etc. To provide a simple example, if we consider an interaction that can be summarized by a Hamiltonian as $H= \Omega \, \sigma^z_i\cdot\sigma^z_j $, that can be obtained from the one in eq.~(\ref{Hamiltoniana_coppia}) assuming $\omega=0$, we can immediately recover from eq.~(\ref{purity}) and eq.~(\ref{probability_interacting_2}) that $\Delta_T=0$ even in the presence of a non-vanishing entanglement.

A simple numerical analysis of the above model shows that, for many mixed particle systems, the non-unitary evolution effect is negligible. However, as we will show in the next section, this effect is a many-body effect and, hence, its relevance increases proportionally to the number of particles in the system.

\section{Oscillation of $N$ interacting neutral particles}
\label{sec_N}

We now generalize the scheme presented above to the case where each copy of the analyzed system is made of a large number $N$ of particles. We consider the same assumptions used in the previous section: 1) we assume that the system is closed; 2) we take into account only the gravitational interaction among the particles; 3) we assume the identity between inertial and gravitational masses; 4) we consider the Newtonian potential valid; 5) we assume the invariance of the relative distances among the fields during the time evolution.

Within the above hypothesis, the system evolves with time-independent Hamiltonian which generalize that in eq.~(\ref{Hamiltoniana_coppia}), i.e.
\begin{eqnarray}
\label{Hamiltoniana_large}
H^{(N)}=\sum_i \omega_i \sigma_i+ \frac{1}{2}\sum_{i,j}\Omega_{i,j} \sigma_i\cdot\sigma_j \, .
\end{eqnarray}
The main difference between eq.~(\ref{Hamiltoniana_coppia}) and eq.~(\ref{Hamiltoniana_large}) is that, now, all parameters of the Hamiltonian depend on the index running on the set of particles. Indeed, $\omega_i$ and $\Omega_{i,j}$ are now given by
$\omega_i=\omega_0+\sum_{j}g_{i,j}(m_1^2-m_2^2)$ and \mbox{$\Omega_{i,j}=g_{i,j}(m_1-m_2)^2$}, where $g_{i,j}=\frac{G}{4d_{i,j}}$ and $d_{i,j}$ is the relative distance between the $i$-th and the $j$-th fields. Despite this loss of symmetry, the Hamiltonian in eq.~(\ref{Hamiltoniana_large}) still holds the fundamental property that it can be seen as a sum of mutual commuting terms. This property plays a key role in the rest of our paper. Indeed, usually the dynamic of a quantum many--body system is extremely complex to be analyzed exactly and numerical and/or approximate approach are needed. However this is not the case. In fact, exploiting such a property, we have that the time evolution operator can be written as the product of several operators each one of them accounts for the evolution induced by a single term of the Hamiltonian in eq.~(\ref{Hamiltoniana_large}). Hence, collecting all the terms it is possible to obtain an exact expression of the state at a time $t>0$ and, more important for our analysis, of its projection into the Hilbert space defined on a single particle.

As in the previous section we consider, at time $t\!=\!0$, that the system is described by a fully separable state. We assume that the first $M$ particles are created in the state $\ket{n_A}$ and the rest is in the state $\ket{n_B}$, so that the initial state is \mbox{$\ket{\psi^{(N)}(0)}=\bigotimes_{\alpha=1}^M\ket{n_A}_\alpha\bigotimes_{\beta=M+1}^{N}\ket{n_B}_\beta$}. Soon after $t\!=\!0$ the system will start to evolve under the influence of the self-gravity and for any $t>0$, the whole system is represented by the pure state (because we are assuming that the system is closed) \mbox{$\ket{\psi^{(N)}(t)}=U(t)\ket{\psi^{(N)}(0)}$}, where the unitary time evolution operator is $U(t)=\exp(-\imath t H^{(N)})$. Knowing the initial state, the reduced density matrix on the selected $k$-th particle can be obtained in terms of the Pauli matrix~\cite{Osborne2002} as
\begin{equation}
\label{reduced_density_general}
\!\rho_{k}(t)\!=\!\frac{1}{2}\!\left( \! \mathbb{1}\!+\!
\sum_{\alpha} \! \bra{\psi^{(N)}(0)}\! U^\dagger\!(t) \sigma^\alpha_k U\!(t) \!\ket{\psi^{(N)}(0)} \sigma^\alpha_k \right)\!\!,\!\! \!\!\!\!\! 
\end{equation}
where $\alpha$ runs over the ensemble $\{x,y,z\}$. Since all terms in the Hamiltonian commutes with each other, the operator $U(t)$ can be arranged as the product of three different terms, i.e. $U(t)=u_k(t)u_{k,r}(t) u_r(t)$. Here \mbox{$u_k(t)=\exp(-\imath \omega_k \sigma^z_k t)$} is the part of the unitary evolution that acts only on the selected $k$-th particle, \mbox{$u_{k,r}(t)=\exp(-\imath t \sum_{j}\Omega_{k,j} \sigma_k\cdot\sigma_j )$} while $u_r(t)$ includes all the other Hamiltonian terms that do not involve directly the $k$-th field. 

Taking into account the fact that Pauli operators on different particles commute with each other, we have that in the evaluation of $\rho_k(t)$ the operator $u_r(t)$ can be neglected. Hence, $\rho_{k}(t)$ depends only on $u_k(t)$ and $u_{k,r}(t)$. Moreover, since both $u_k(t)$ and $u_{k,r}(t)$ depend only on $\sigma^z_k$, we have that the coefficient of $\sigma_k^z$, and, hence, the elements on the diagonal of the reduced density matrix are time-independent. On the contrary, the coefficients of $\sigma_k^x$ and $\sigma_k^y$ depend on time and their derivation is long but straightforward. Substituting the expression of the flavor fields in eq.~(\ref{leptonic_state}) in $\ket{\psi(0)}$, we have that the initial state can be written as
\begin{equation}
\label{initial_state_rho_k}
\ket{\psi(0)}=\sum_{\{l\}} R_l \left(a_k\ket{m_1,l}+b_k \ket{m_2,l}\right),
\end{equation}
where $\ket{m_1,l}$ $(\ket{m_2,l})$ is a generic tensor product of mass states in which the state in the $k$-th field is equal to $m_1$ ($m_2$). For the different parameters, $a_k$ ($b_k$) is equal to $\cos(\theta)$ $(e^{\imath \phi} \sin(\theta))$ for $k\le M$ and to $e^{-\imath \phi} \sin(\theta)$ $(\cos(\theta))$ for $k>M$. On the other hand, $R_l=\prod_s c_{l,s}$, where $c_{l,s}$ is equal to $a_l$ ($b_l$) if in $\ket{l}$ the $s$-th field in the mass state $m_1$ ($m_2$). 

From this expression, it is immediate to obtain the expression of $\ket{\tilde{\psi}(t)}=u_k(t) u_{k,r}(t)\ket{\psi(0)}$,
\begin{equation}
\label{initial_state_rho_k_2}
\!\ket{\tilde{\psi}(t)}\!=\!\sum_{\{l\}}\! R_l \!\left(\!a_k e^{-\imath (\omega+\Gamma_k) t}\ket{1,l}+b_k e^{\imath (\omega+\Gamma_k) t} \ket{2,l}\!\right),\!\!\!
\end{equation}
where $\Gamma_k=\sum_{s}(-1)^{\lambda_{s}}\Omega_{s,k}$ with $\lambda_s=1$ ($\lambda_s=2$) if in $\ket{l}$ the mass state of the $s$-th particle is $m_1$ $(m_2)$.

The knowledge of $\ket{\tilde{\psi}(t)}$ allows us to construct the reduced density matrix taking into account that $\bra{\psi^{(N)}(0)} U^\dagger(t) \sigma^\alpha_k U(t) \ket{\psi^{(N)}(0)}=\bra{\tilde{\psi}(t)}\sigma^\alpha_k\ket{\tilde{\psi}(t)}$. After some algebras we obtain the following general exact expression for the reduced density matrix,
\begin{equation}
\label{reduced_density}
\!\rho_{k}(t) \!=
\!\frac{1}{2}\!\left(
\begin{array}{cc}
1 +\zeta_k \cos(2 \theta) & \zeta_k e^{-\imath \phi} \sin(2 \theta ) a_k^*(t) \\
\zeta_k e^{\imath \phi} \sin(2 \theta ) a_k(t) & 1 -\zeta_k \cos(2 \theta)
\end{array}
\right),\!\!\!\!\!\!\!
\end{equation}
where $\zeta_k$ is a function that is equal to $+1$ for $k\le M$, and to $-1$ for $k > M$, $a_k(t)$ is given by
\begin{equation}
\label{definition_of_a}
 a_k(t)\! =\! e^{\imath 2 \omega_k t }\!\prod_{j=1}^N\!(\cos(2 \Omega_{k,j}t)\!+\!\imath \zeta_k\cos(2\theta)\sin(2 \Omega_{k,j}t)),\!\!\!\!
\end{equation}
and we assume, as definition, that $\Omega_{k,k}=0$. It is worth to note that the expression of the reduced density matrix in eq.~(\ref{reduced_density}) is exact and obtained, once given the set of the relative distances, without any approximation and without the necessity to use any master equation approach. 

As we have already said, since we are neglecting any interaction among the elements of the system and the surrounding world, the time evolution is unitary. As a consequence, $\ket{\psi^{(N)}(t)}$ is always a pure state. Therefore, it is possible to use the $2$--Renyi entropy defined as $S_2=-\ln(\mathcal{P}(\rho_i(t)))$ to quantify the total entanglement that any single particle shares with the rest of the system. From eq.~(\ref{reduced_density}), we obtain for the time-dependent purity
\begin{equation}
\label{purity_many}
\mathcal{P}(\rho_k(t))=1-\sin^2(2 \theta) \left(1-|a_k(t)|^2\right) \,.
\end{equation}
Now the couplings $\Omega_{k,j}$ are not invariant under the change of fields. Thus, we have $|a_k(t)|^2<1, \, \forall t>0$, which reduces to 1 only at $t=0$. Therefore, for $t>0$, any single particle is entangled with the rest of the system.

By means of eq.~(\ref{reduced_density}) we can generalize the result presented in eqs.~(\ref{probability_interacting}) and eq.~(\ref{probability_interacting_2}) by analyzing the oscillation probability in two copies of the system in which the first one has $M=N$ and the second one has $M=0$. Differently from the previous case, now the reduced density matrices at $t>0$, and, hence, also the oscillation probabilities, are site-dependent. Thus, we have to consider the average over all elements of the system. Explicitly we obtain
\begin{eqnarray}
\label{probability_interacting_many}
P_{n_A \rightarrow n_B} \! &\!=\!& \frac{1}{2} \sin^2(2\theta)\left(1 - \frac{1}{N}\!\sum_{k=1}^N \mathrm{Re}(a_k^{(A)}(t))\right) \,;
\nonumber \\
P_{n_B \rightarrow n_A} \! &\!=\!& \frac{1}{2} \sin^2(2\theta)\left(1 - \frac{1}{N}\!\sum_{k=1}^N \mathrm{Re}(a_k^{(B)}(t))\right)\,,
\end{eqnarray}
where $\mathrm{Re}(a_k^{(A)}(t))$ ($\mathrm{Re}(a_k^{(B)}(t))$) is the real part of $a_k^{(A)}(t)$ ($a_k^{(B)}(t)$) that are the functions $a_k(t)$ when $M=N$ ($M=0$). As well as in eq.~(\ref{probability_interacting}), also the transition probability in eq.~(\ref{probability_interacting_many}) does not depend on the $CP$-violating Majorana phase, so that $\Delta_{CP} = 0$. On the other hand, $P_{n_A \rightarrow n_B} \neq P_{n_B \rightarrow n_A}$ because of $a_k^{(A)}(t)\neq a_k^{(B)}(t)$ and, hence, $\Delta_T\neq 0$. In order to make this violation more evident, let us assume that $\Omega_{k,j} t\ll 1$. In this case $a_k(t)$ becomes
\begin{equation}
\label{approx_a}
\!\!a_k(t)\!\simeq\!e^{\imath 2 \omega_k t }
\left(1 \pm 2 \imath \cos(2 \theta) \sum_{j=1}^N \Omega_{k,j}t \right)\,,
\end{equation}
where the sign $+$ is for the system composed at $t=0$ only of $n_A$-particles, i.e. $a_k^{(A)}(t)$, and the sign $-$ is for systems composed at $t=0$ only of $n_B$-particles, i.e. $a_k^{(B)}(t)$. Substituting eq.~(\ref{approx_a}) in eqs.~(\ref{probability_interacting_many}), we have the time-reversal symmetry violation becomes
\begin{eqnarray}
\label{probability_interacting_many_2}\nonumber
\!\!\!\!\!\!\!\Delta_T & = & P_{n_A \rightarrow n_B} - P_{n_B \rightarrow n_A}
\\ & = &
\sin^2(2 \theta) \cos(2\theta) \frac{2 t}{N}\!\! \sum_{k,j=1}^N \sin(2 \omega_k t) \Omega_{k,j}\,.
\end{eqnarray}
Since $\Delta_{CP} \neq \Delta_T$, the $CPT$-symmetry is broken.

The exact value of the violation of the time-reversal symmetry in eq.~(\ref{probability_interacting_many_2})depends on the whole set of relative distances among the particles in the system. For system with large $N$ this set is not known but we can express $\Delta_T$ in terms of average values. Indeed, since the gravity has a very long range, then the sum in eq.~(\ref{probability_interacting_many_2}) contains $N(N-1)$ non--zero terms (we assumed $\Omega_{i,i}=0 \; \forall i$). Moreover, all these terms have the same sign. In fact, since gravity is attractive, all $\Omega_{k,j}$ are negative regardless of the particular choice of $k$ and $j$. On the other hand, the sign of $\omega_{k}$ depends on the difference between $m_1$ and $m_2$ and, hence, the sign does not depend on $k$. Therefore, inside the sum, for time short enough such that $\max(\omega_k t)<\pi/4$, all the terms have the same sign. In other words, by defining $f_k=\frac{\sin(2 \omega_k t)}{N}\sum_j \Omega_{k,j}$ we have that $f_k$ is of the order of unity and all $f_k \, \forall k$ have the same sign. Therefore, defining $F$ as the average of $f_k$, i.e. $F=\frac{1}{N}\sum_{k=1}^N f_k$ we have 
\begin{eqnarray}
\label{probability_interacting_many_3}
\Delta_T & = & 
\sin^2(2 \theta) \cos(2\theta) 2 N t F\, ,
\end{eqnarray}
where we see explicitly that $\Delta_T$ is proportional to the number of particles of the system.

Similar $CPT$ violation can be obtained for all configurations in which the difference $M$ and $N-M$ is of the same order of magnitude of $N$. Indeed, when this does not happen, as in the case in which at $t=0$ we have $N/2$ particles in the flavor state $\ket{n_A}$ and $N/2$ in $\ket{n_B}$, it is possible to show, using the Lindeberg--L\'evy theorem~\cite{VanderVaart1998}, that $\Delta_T$ is proportional to $\sqrt{N}$.

\section{Time-dependent relative distances}

All the results obtained up to now were derived assuming constant the relative distances among the particles of the system. However, this assumption is not crucial. In this section we extend our analysis also to the more realistic case in which the distances change in time. 

Removing the constraints of the independence of the distances on time, for any $t>0$, the Hamiltonian of the gravitational self-interacting system can be written as 
\begin{eqnarray}
\label{Hamiltoniana_large_time}
H^{(N)}(t)=\sum_k \omega_k(t) \sigma_k+ \frac{1}{2}\sum_{k,j}\Omega_{k,j}(t) \, \sigma_k\cdot\sigma_j \, .
\end{eqnarray}
Differently from eq.~(\ref{Hamiltoniana_large}), now the parameters $\omega_k(t)$ and $\Omega_{k,j}(t)$ depend explicitly on time. Indeed, they are, respectively, $\omega_k(t)=\omega_0+\sum_{j}g_{k,j}(t)(m_1^2-m_2^2)$ and \mbox{$\Omega_{k,j}(t)=g_{k,j}(t)(m_1-m_2)^2$}, where $g_{k,j}(t)=-\frac{G}{4d_{k,j}(t)}$ and $d_{k,j}(t)$ is the relative time-dependent distance between the $k$-th and the $j$-th fields.

The time dependence of the Hamiltonian affects the evaluation of the time evolution unitary operator. Indeed, this operator, that is obtained as solution of the Schr\"{o}dinger equation can be, in general, written in terms of Magnus expansion~\cite{Blanes2009,Blanes2010} as
\begin{equation}
 \label{magnus_1}
 U(t)=\exp\left(\sum_l \frac{1}{l!}\Lambda_l\right),
\end{equation}
where the first terms of the expansion are 
\begin{eqnarray}
 \label{magnus_2}
 \Lambda_1&=& -\imath \int_0^t H^{(N)}(\tau) d\tau; \\
 \Lambda_2&=& (-\imath)^2 \int_0^t \int_0^{\tau_1} [H^{(N)}(\tau_1),H^{(N)}(\tau_2)] d\tau_1 d\tau_2, \nonumber
\end{eqnarray}
where the square brackets denote the commutator between the Hamiltonian at diffferent times. Moreover, all the other terms of the Magnus expansion depend on a combination of commutators between the Hamiltonian in eq.~(\ref{Hamiltoniana_large_time}) at different times~\cite{Blanes2010}. 

However, in the case that we are analyzing, it is easy to check that \mbox{$[H(\tau_1),H(\tau_2)]=0$} $\forall\tau_1,\,\tau_2 $ and, hence, we have that $\Lambda_l=0\, \forall\, l \ge 2$. Therefore, from eq.~(\ref{magnus_1}) we obtain that 
\begin{eqnarray}
 \label{magnus_3}
 U(t)\!&\!=\!&\!\!\exp\left(-\imath\! \int_0^t \! H(\tau) d\tau \right) \nonumber \\ 
 \!&\!=\!&\!\!\exp\left[\!-\imath \!\! \int_0^t \!\!\left(\sum_k \omega_k(\tau) \sigma_k \!+\!\sum_{k,j}\frac{\Omega_{k,j}(t)}{2} \sigma_k \cdot\sigma_j \right) \! d\tau\! \right]\nonumber \\
 \!&\!=\!&\!\!\exp\left(-\imath t \sum_k \tilde{\omega}_k(t) \sigma_k \!-\frac{\imath t}{2} \!\sum_{k,j}\tilde{\Omega}_{k,j}(t) \sigma_k \cdot \sigma_j \right)\nonumber \\
 \!&\!=\!&\!\!\exp\left(-\imath t \tilde{H}(t)\right),
\end{eqnarray}
where
\begin{eqnarray}
 \label{replacing_large}
 \tilde{H}(t)& = &\sum_k \tilde{\omega}_k(t) \sigma_k \!+\frac{1}{2} \!\sum_{k,j}\tilde{\Omega}_{k,j}(t) \sigma_k \cdot \sigma_j\, ; \nonumber \\
 \tilde\omega_k &=& \omega_0 - \frac{1}{t} \frac{G}{4}(m_1^2-m_2^2) \sum_j \int_0^t \frac{1}{d_{k,j}(\tau)} \mathrm{d}\tau \nonumber \,;\\
 \tilde{\Omega}_{k,j} &=& -\frac{1}{t} \int_0^t \frac{G}{4 d_{k,j}(\tau)}(m_1-m_2)^2 \mathrm{d}\tau \,.
\end{eqnarray}
The time evolution operator in eq.~(\ref{magnus_3}) is, formally, equivalent to that obtained in Sec.~\ref{sec_N}. Therefore, independently on $N$ and $M$, we can use the relations in eq.~(\ref{replacing_large}) to generalize the results obtained in Sec.~\ref{sec_N} to the case of time-dependent relative distances.

Before to conclude, it is worth to underline that this surprising result holds only because \mbox{$[H(\tau_1),H(\tau_2)]=0$} $\forall\tau_1,\,\tau_2 $. In the general case this is not true and, hence, the generalization to time-dependent relative distances cannot be evaluated exactly. In these cases we are forced to use different approaches, such as master equations, Dyson series expansions, etc.

\section{Conclusions}

We have shown that the gravity in a self-interacting particles mixing system leads to the $CPT$ violation. This violation is related to the emergence of a non-zero entanglement among the elements of the system induced by a difference of mass of the free fields. Moreover, since the gravitational interaction is additive, the $CPT$ violation is proportional to the number of elements of the system and its density. Therefore, this effect could play a crucial role in galactic objects and in the first stage of the Universe where the densities and the number of particles are very high.  

The $CPT$-symmetry violation presented in this paper is not the first one discovered in the context of neutral particle oscillations. In neutrino physics, several studies have been devoted to the analysis of symmetry violations induced by dissipative dynamics~\cite{Benatti2000,Benatti2001,Capolupo2018}. However, our work presents several aspects of novelty. In fact, instead to consider a single particle as an open system affected by several uncontrolled phenomena, we consider an ensemble of self-interacting particles as a closed system with all the physical quantities being under control. As a consequence we have a difference $CPT$ violation. In fact, in the previous works, $CPT$ violation was generated by a $CP$-symmetry breaking and not, as in our case, by a violation of the $T$-symmetry.

However, our results must not be considered in contrast with the ones presented in Refs.~\cite{Benatti2000,Benatti2001,Capolupo2018}. Indeed, the non-unitary dynamics includes a wide family of physical sources of decoherence. On the contrary, we have limited our analysis to the effects due to the self-gravitational interaction so neglecting all other possible sources of decoherence~\cite{Kostelecky1995,Kostelecky2004,Colladay1997,Amelino-Camelia2010,Bernabeu2017}. Nevertheless, our work paves the way to several other works, in which a detailed analysis of each individual contribution to decoherence can be realized.

Moreover, it is worth to note that the mechanism here presented is not only limited to gravitational interaction. In fact, the two main requirements are: 1) the presence of neutral particles whose flavor states are superpositions of the eigenstates of a free field Hamiltonian; 2) The presence of an interaction depending on the eigenstates of the free Hamiltonian. Within this hypothesis the interaction, not necessarily of gravitational origin, among two or more of these particles will generate entanglement and, hence, induce a $CPT$-symmetry violation in the flavors oscillations.

Table-top experiments, based on Rydberg atoms confined in microtraps and optically manipulated~\cite{Pillet2016}, can simulate the mixing and the mutual interaction. In this system, the two internal states, i.e. the ground state and the excited Rydberg level can represent the mass eigenstates, whereas two particular orthonormal superpositions can simulate the flavor states, and the dipole-dipole interaction can play the role of the gravity~\cite{Jau2016,Labuhn2016,Ravets2014}. Thus, next experiments on atomic physics could allow to test the fundamental laws and symmetries of nature.

\section*{Acknowledgements}

K.S. is grateful to the Austrian Science Fund (FWF-P30821, FWF-P26783). A.C. thanks partial financial support from MIUR and INFN and the COST Action CA1511 Cosmology and Astrophysics Network for Theoretical Advances and Training Actions (CANTATA). SMG acknowledges support from the European Regional Development Fund for the Competitiveness and Cohesion Operational Programme (KK.01.1.1.06--RBI TWIN SIN) and from the Croatian Science Fund Project No. IP-2016--6--3347. SMG also acknowledges the QuantiXLie Center of Excellence, a project co--financed by the Croatian Government and European Union through the European Regional Development Fund--the Competitiveness and Cohesion Operational Programme (Grant KK.01.1.1.01.0004).

\end{document}